\documentclass[useAMS,usenatbib]{mn2e}
\usepackage[dvips]{graphicx}

\usepackage{amsmath}
\usepackage{bm}
\usepackage{natbib}
\usepackage{dcolumn}

\newcommand{\be}{\begin{equation}}
\newcommand{\ee}{\end{equation}}
\newcommand{\bea}{\begin{eqnarray}}
\newcommand{\eea}{\end{eqnarray}}
\title[More on Lensing by a Cosmological Constant]{More on Lensing by a Cosmological Constant}
\author[M. Ishak, W. Rindler, J. Dossett]{M. Ishak\thanks{E-mail:
mishak@utdallas.edu}, W. Rindler\thanks{F.R.A.S.}, J. Dossett\\ \\
Department of Physics, The University of Texas at Dallas, Richardson, TX 75083, USA}

\begin{document}

\date{\today}

\pagerange{\pageref{firstpage}--\pageref{lastpage}} \pubyear{0000}

\maketitle

\label{firstpage}

\begin{abstract}
The question of whether or not the cosmological constant affects the 
bending of light around a concentrated mass has been the subject of 
some recent papers.  We present here a simple, specific and transparent 
example where $\Lambda$ bending clearly takes place, and where it is 
clearly neither a coordinate effect nor an aberration effect.  
We then show that in some recent works using perturbation theory 
the $\Lambda$ contribution was missed because of initial too-stringent 
smallness assumptions. Namely:  Our method has been to insert a Kottler 
(Schwarzschild with $\Lambda$) vacuole into a Friedmann universe, 
and to calculate the total bending within the vacuole.  We assume that 
no more bending occurs outside. It is important to observe that while 
the mass contribution to the bending takes place mainly quite near the 
lens, the $\Lambda$ bending continues throughout the vacuole. Thus if 
one deliberately restricts one's search for $\Lambda$ bending to the 
immediate neighborhood of the lens, one will not find it. Lastly, we 
show that the $\Lambda$ bending also follows from standard Weyl focusing, 
and so again, it cannot be a coordinate effect.
\end{abstract}

\begin{keywords}
gravitational lensing -- gravitation -- cosmology: theory.
\end{keywords}

\section{introduction and example}

Since the publication in 2007 of our first paper on this subject \cite{RindlerAndIshak2007}, in which we claimed that $\Lambda$ contributed to the bending of light, a number of authors have supported our findings \cite{Schucker1,Lake2007,Schucker2,Sereno2007,Schucker3,Miraghaei} whilst others have questioned them \cite{Park,SPF,Khriplovich}. Though perhaps not of immediate practicality, the question of whether $\Lambda$ does or does not contribute to the bending of light around a concentrated mass is of theoretical interest and deserves to be settled.  In the present paper we hope to contribute to this end.

Very early on, Eddington addressed the related problem of whether $\Lambda$ affects the orbits of particles around a concentrated mass and found that indeed it does; it adds a minute correction to the advance of the perihelion of the planets [Eq. (14.25) and Ex. 14.8 in \cite{Rindler}]. Eddington apparently did not investigate the orbits of photons in this regard.

Indeed, the usual orbit equations for particles degenerate in the case of photons (i.e., when $ds^2 = 0$) in that the $\Lambda$ term drops out, see Eq.(7) in \cite{RindlerAndIshak2007}.  However, it is our contention that $\Lambda$ nevertheless affects the photon orbits because the $\Lambda$-independent coordinate equations now refer to a $\Lambda$-dependent geometry.

In this first section we begin by exhibiting a specific simple and transparent  example,  where $\Lambda$ bending clearly takes place, and which in itself should suffice to answer those of our critics who categorically deny that $\Lambda$ contributes to the bending of light.  The succeeding sections below deal with more extensive topics that have been the subject of debate.

The example in question is a simple consequence of results contained already in our original paper \cite{RindlerAndIshak2007}. There is a well-known generalization of the Schwarzschild metric that includes $\Lambda$. It is called the Kottler \cite{Kottler} metric and represents the unique solution of Einstein's vacuum field equations with Lambda for the spacetime around a spherically symmetric mass.  By an extension of Birkhoff's theorem, it remains valid in the vacuum between the central mass and any spherically symmetric (not necessarily static) mass distribution around it, such as a Friedmann universe. It looks as follows:
\begin{equation}
ds^2=\alpha(r) dt^2 - \alpha(r)^{-1} dr^2-r^2 (d\theta^2+sin^2(\theta) d\phi^2),
\label{eq:metric}
\end{equation}
where relativistic units are used $(c = G = 1)$ and
\begin{equation}
\alpha(r) \equiv 1-\frac{2m}{r}-\frac{\Lambda r^2}{3}.
\label{eq:alpha}
\end{equation}
For small $r$ it approximates Schwarzschild space, and for large $r$,  if $\Lambda>0$, de Sitter space (in static form). For the sake of our present simplified example, we assume that this de Sitter space is the actual universe.

Now consider a ray of light in the coordinate equatorial plane  $\theta=\pi/2$, bending closely around the central mass (the "lens") and connecting two diametrically opposite points P and Q equidistant from the center ($r_P = r_Q$,  $\phi_P = 0$, $\phi_Q = \pi$).  (See Fig.2 of \cite{RindlerAndIshak2007}) If Q coincides with a luminous source, and P with an inertial observer momentarily at rest in the metric, the observer will measure an angle
\begin{equation}
\psi_0 \approx \frac{2m}{R} -\frac{\Lambda R^3}{12 m}
\label{eq:result1}
\end{equation}
between that ray and a ray from the center of the lens (R being the radius of the lens).  This formula (to first order in $m/R$ and $\Lambda R^3/m$) was established in Eq. (17) of \cite{RindlerAndIshak2007}, and clearly shows how $\Lambda$ affects the bending.  A whole cone of such rays will, in fact, reach the observer from all around the lens, who therefore sees the source spread out into an "Einstein ring" of angular diameter $2\psi_0$. 

Of course, our stationary observer is not a "cosmological" observer moving with the Hubble flow in the de Sitter universe.  Such a cosmological observer has a local Hubble velocity  
\be
V \approx H\,r = \sqrt{\Lambda/3}\,r
\ee
 and an aberration factor
\be 
\left(\frac{1+V}{1-V}\right)^{1/2}
\ee
 relative to the stationary observer (Eq. (4.9) in \cite{Rindler}): his Einstein ring appears that much larger. Of course, this factor trivially brings another $\Lambda$ contribution into play, but this is NOT the one that is under scrutiny here. It does not affect the RATIO in which $m$ and $\Lambda$ contribute to the bending.

    In Simpson et al. \cite{SPF}, Sec.IV, it is suggested that our claimed Lambda bending is
really an aberration effect, and earlier, in Sec II, that it is a gauge
effect:  ``The potential-like term  $\Lambda r^2/3$  in the Kottler  $f(r)$ [our
$\alpha(r)$] ... does not arise from the true potential $\Phi$, and thus it should
not be taken to cause lensing ... it seems fair to assert that the
appearance of a lensing effect from $\Lambda$ in the Ishak-Rindler analysis is
purely a gauge artefact."  Our example above suggests otherwise. It
dispenses with the potential formalism of the usual lensing literature, and
instead deals directly with the spacetime and its null geodesics in a
geometric (coordinate-independent) way.  So the question of a "true
potential" or of a gauge choice never arises.
\section{The $\Lambda$ contribution to light bending and approximations in the Newtonian Gauge}

In this section we discuss how the argument presented in Simpson et al. \cite{SPF} is based on an assumption that eliminates the $\Lambda$ contribution to the bending of light {\textit {a priori}}. As the authors acknowledge after equation (9) of their section II and at the end of that section, it was assumed in their work that in a perturbed FLRW universe the Kottler vacuole around the lens is negligibly small in comparison with the Hubble length and also that, in most of the volume of the vacuole the radius $r$ is almost of the same magnitude as the proper radius of the vacuole. 

First, as discussed in our previous work \cite{Ishaketal2008,Ishak2008}, the radius of the vacuole, $r_b$, is much larger than the radius $R$ of the lens itself and when using any expression with the radii squared then the corresponding ratio is $(\frac{r_b}{R})^2\sim 10^3$, for example, for the cluster Abel2744 \cite{Smail1991,Allen}, this ratio is $\left(\frac{r_b}{R}\right)^2\approx 2500$. 

Second, the ratio of the size of the vacuole compared to the Hubble radius is about the same order of magnitude as the ratio between the $\Lambda$-contribution to the bending angle compared to the first-order Einstein angle. 
Indeed, we calculated in \cite{Ishaketal2008}, for several clusters, the contribution of the $\Lambda$ term to the bending angle and found it equal to about $10^{-3}-10^{-4}$ of the value of the first-order Einstein angle. And as we discussed there, while this is small, it is as big as the second-order term in the mass and in some cases even larger. Now, the diameter of the vacuole around the clusters is about $d_{vacuole}\approx 2-10\,Mpc$, (for example $5\,Mps$ for Abel2744), so the ratios between $d_{vacuole}$ and the angular-diameter distances involved in the lens equation are of the order $10^{-2} - 10^{-3}$ and finally the ratio $d_{vacuole}/R_{Hubble}\approx 10^{-3}-10^{-4}$. So by neglecting the size of vacuole, the effect of $\Lambda$ has been neglected with it as we show below. 

Another relevant point here is that the bending due to the mass is achieved quickly in the vicinity of the lens and then changes very slightly as we move away from the lens, whereas the effect of $\Lambda$ inside the vacuole accumulates up to the boundary of the vacuole. This makes the larger size of $r_b$ significant.  
  
We next show how not discarding the diameter of the vacuole compared to the Hubble radius affects the calculation of \cite{SPF} and restores the contribution of $\Lambda$ to the bending angle.

Ref. \cite{SPF} considered the perturbed Robertson-Walker metric where scalar metric fluctuations are described by scalar potentials, $\Phi$ and $\Psi$:
\[
ds^2 = (1+2\Phi)\, dt^2 - a^2(t)\, (1-2\Psi)\, (d\chi^2 + \chi^2
d\Omega^2)\label{eq:FRW}
\]
and where $\chi$ is the comoving radius, and $d\Omega$
the element for the unit sphere. The case of a flat universe with no anisotropic stresses was considered, so $\Psi=\Phi$. And then, the question of how the perturbed FRW metric compares with the exact Kottler metric inside the vacuole was explored. After a coordinate transformation and some approximations, the authors of \cite {SPF} arrived at the following expression for their function $f$ [in their equation (9)]:  
\bea
f = 1 -2\chi\Phi' - \dot a^2\chi^2 \left[1-2\Phi 
\left(\frac{\partial \ln |\Phi|}{\partial \ln a}+2\right)\right].
\label{eq:f_function}
\eea
At this point, we don't make the same simplification as was made in \cite{SPF} 
 and we don't ignore the terms proportional to $\Phi$ in the square brackets 
since we want to keep the contribution from the vacuole even if it is small 
compared with the Hubble length. 

Next, we recall the $00$-equation of the perturbed Einstein equations (see for example equation (5.27) in \cite{Dodelson}):  
\be
k^2 \Phi +3 \frac{\dot{a}}{a}\left(\dot\Phi-\Phi \frac{\dot a}{a}\right)= 4 \pi G a^2 \rho_m \delta_m.
\label{eq:poisson}
\ee
At the same level of approximation where we want to keep the contribution from the vacuole, we don't drop the higher order terms on the LHS of (\ref{eq:poisson}) and we don't use the approximation ${\partial \ln |\Phi|}/{\partial \ln a}=-1$ also invoked in \cite{SPF}. One could for example model small departures from this relation by $-1\pm \epsilon$.    

Now, using (\ref{eq:f_function}), (\ref{eq:poisson}), the Friedmann equation, and following the same procedure as in section II of \cite{SPF}, we integrate the potential $\Phi$ and obtain the solution 
\be
\Phi = -{m \over r} - {mr^2 \over 2 r_b^3} - {{\Lambda r^2}\over 12}
\label{eq:phi}
\ee
plus smaller terms. 

Remarkably, our integration of the potential gives the 2 mass terms of reference \cite{SPF} plus $\Lambda$ terms with a leading term that is precisely $-\frac{\Lambda r^2}{12}$ \footnote{We verified that when our full solution for $\phi$ (with 3 mass terms plus 4 terms with $\Lambda$) is put into equation (\ref{eq:f_function}), it yields the function $\alpha(r)$ in (\ref{eq:alpha}) as it should plus higher order terms, i.e. $\rm{O}(m\Lambda)$.
}.

This $\Lambda$ term in the potential gives the same $\Lambda$ contribution to the light-bending angle (see for example \cite{Ishak2008}) as the one derived from different approaches in \cite{RindlerAndIshak2007,Ishaketal2008,Ishak2008}, namely,
\be
\alpha_{\Lambda}=-\frac{\Lambda\,R\,r_b}{3}
\label{eq:LambdaBending}
\ee
where we assumed $R\ll r_b$ \footnote{Indeed our expression for $\alpha_{\Lambda}$ was derived from equations (18) in \cite{Ishaketal2008} where we assumed a small angle $\Phi_b=\frac{R}{r_b}$. Similarly, we justifiably assumed in the step from equation (11) to (12) in \cite{Ishak2008} that $\big{(}\frac{R}{r_b}\big{)}^2\ll1$ thus reducing $\alpha_{\Lambda}=-\frac{\Lambda\,R\,r_b}{3}\sqrt{1-\big{(}\frac{R}{r_b}\big{)}^2}$ to simply $\alpha_{\Lambda}=-\frac{\Lambda\,R\,r_b}{3}$.
}.

The result (\ref{eq:phi}) above for the potential shows that when the size of the vacuole is not neglected in comparison with the Hubble length then the potential does contain $\Lambda$-terms (as found previously \cite{Ishaketal2008,Ishak2008}) and confirms that $\Lambda$ contributes to the bending of light inside the vacuole. 

Therefore, we find that the $\Lambda$ contribution terms were missed in \cite{SPF} because of initial too-stringent smallness assumptions on the size of the vacuole.  

It was also stated in the same paper by Simpson et al. \cite{SPF} (section IV) that the $\Lambda$ bending term in our equation (\ref{eq:result1}) (see \cite{RindlerAndIshak2007}) and equation (\ref{eq:LambdaBending}) above (see \cite{Ishaketal2008}) appear problematic since they do not vanish when the lensing mass is taken to be zero. Clearly when the lens mass is zero (i.e. no lens), the radius of the lens is zero (i.e. $R=0$) and the $\Lambda$-contribution term vanishes as well (i.e. $\alpha_{\Lambda}=0$). So, as expected, when there is no lensing mass inside the de Sitter vacuole, $\Lambda$ has no effect on the bending of light but when a central mass is present as in the Kottler space that is under consideration here, the mass bends the ray of light while $\Lambda$ diminishes the deflection inside the vacuole. 

Finally, it is worth noting that the structure of the Weyl tensor (see Eq. (\ref{eq:WeylTensor}) in the next section) for the Kottler spacetime shows that inside the Kottler vacuole, the Weyl tensor is nonvanishing and so the spacetime is not isotropic as was assumed in \cite{SPF} after equation (22). This invalidates the argument used after equation (22) there.  
\\
\section{The contribution of $\Lambda$ to the bending of light from the Weyl Focusing}
The propagation of light in a given space-time can be rigorously described using the theory of geometrical optics in General Relativity \cite{Sachs1961,Dyer1977}. The distortions of light bundles are described by the optical scalar equations driven by Ricci and Weyl focusing \cite{Sachs1961,Dyer1977,Schneider1992}. We recall that the Ricci focusing is given by \cite{Sachs1961,Dyer1977,Schneider1992}
\begin{equation}
\mathcal{R}=R_{ab}k^{a}k^{b}
\end{equation} 
and the Weyl focusing (up to a phase factor) by
\begin{equation}
\mathcal{F}=C_{aibj}k^{a}k^{b}t^{i}t^{j},
\label{eq:Weyl}
\end{equation} 
where $R_{ab}$ is the Ricci tensor, $C_{aibj}$ is the Weyl tensor, and $k^a$ and $t^{i}$ are null vectors. We here follow the notation of \cite{Dyer1977,Schneider1992}. 

In the Kottler vacuum space ($R_{ab}=0$), the mass and $\Lambda$ contributions to light-bending come from the Weyl focusing as covariantly defined in (\ref{eq:Weyl}). The non-vanishing components of the Weyl tensor read
\bea
C_{t r t r}&=&-\frac{2m}{r} \nonumber \\
C_{t \theta t \theta}&=&\frac{m}{r}(1-\frac{2m}{r}-\frac{\Lambda r^2}{3})=C_{t \phi t \phi}/\sin^2(\theta) \nonumber \\
C_{r \theta r \theta}&=&\frac{-m/r}{1-2m/r-\frac{\Lambda r^2}{3}}=C_{r \phi r \phi}/ \sin^2(\theta) \nonumber \\
C_{\theta \phi \theta \phi}&=& 2\, r\, m\, \sin^2(\theta)
\label{eq:WeylTensor}
\eea
Following \cite{Dyer1977}, it is easy to verify for the Kottler metric (\ref{eq:metric}) that 
\begin{equation}
\mathcal{R}=0\,\,and\,\,\mathcal{F}=\frac{3 m h^2}{r^5} 
\label{eq:RF}
\end{equation}
(equations (27) and (35) in \cite{Dyer1977}). In Schwarzschild space, $h$ is the impact parameter but in Kottler space (not asymptotically flat) $h$ is the constant of motion $J/E$ where $J$ and $E$ are respectively the momentum and the energy of the photon. The relation between the point of closest approach $r_0$ and the constant of the motion $h$ (or $b$ in other notations) is given by (e.g \cite{Wald1984,Kagramanova})
\begin{equation}
\frac{r_0^2}{h^2}=1-\frac{2m}{r_0}-\frac{\Lambda r_0^2}{3}.
\label{eq:r_0toh}
\end{equation}
For Schwarzschild space, the solution $r_0(h)$ of this relation is given on page 145 of \cite{Wald1984}. The solution can be immediately expanded to include $\Lambda$ and is given by 
\begin{equation}
r_0 = \frac{2}{\sqrt{\frac{3}{h^2}+\Lambda}}\cos \Big{(} \frac{1}{3} \arccos \Big{(} -3m \sqrt{\frac{3}{h^2}+\Lambda} \Big{)} \Big{)}, 
\label{eq:r_0}
\end{equation}
in agreement with footnote 9 in \cite{Finelli}. For $\Lambda=0$ it reduces to equation (6.3.37) that was derived for Schwarzschild in \cite{Wald1984}.

Now, if one chooses to express the Weyl focusing fully in terms of the point of closest approach $r_0$ then  (\ref{eq:r_0toh}) and (\ref{eq:RF}) yield  

\begin{equation}
\mathcal{F}=\frac{3 m h^2}{r_0^5} = \frac{3 m}{r_0^3} \Big{(} 1- \frac{2m}{r_0}- \frac {\Lambda r_0^2}{3}  \Big{)}^{-1}. 
\end{equation}

So $\Lambda$ is present in the final result. Similarly, if one chooses to express the focusing fully in terms of $h$, then using (\ref{eq:r_0}) in order to eliminate $r_0$ from (\ref{eq:RF}) gives $\mathcal{F}$ as a function of $m$, $h$, and $\Lambda$. Again the final expression for the Weyl focusing has $\Lambda$ in it. 

This result is independent of angle calculations and supports the conclusion that $\Lambda$ contributes to the bending of light the Kottler space in an invariant and thus eminently coordinate-independent way. 

\section{Further remarks on the contribution of $\Lambda$ to light bending}

Schucker performed calculations in the Kottler space \cite{Schucker1,Schucker2} and also used a piece-wise integration of null geodesics \cite{Schucker3} in the Einstein-Strauss model. He confirmed our result \cite{RindlerAndIshak2007} that a positive cosmological constant decreases the bending of light by an isolated spherical mass. He finds in \cite{Schucker3} that the decrease in the deflection angle due to $\Lambda$ is attenuated by a homogeneous Friedmann background when added around the spherical mass and by the recession of the observer without however canceling it. A result consistent with our work \cite{Ishaketal2008,Ishak2008}. 

Sereno did another calculation \cite{Sereno2007} in the Kottler space and re-derived our $\Lambda$ contribution term, however, he argued that the term can be embedded into the angular diameter distances. In a further analysis \cite{Sereno2007}, he derived another $\Lambda$ contribution term that he argued cannot be included into the angular diameter distances.     

In the work of Khriplovich \& Pomeransky \cite{Khriplovich} on this question, the authors did an expansion of their equation (16) keeping only the first term in the invariant considered there and dropping the second term of that equation (i.e. $\omega \lambda \rho$ in their notation) that contains a $\Lambda$ contribution. They then claim that all factors dependent on $\Lambda$ cancel. Then, the authors do reconsider their position and state in the penultimate paragraph of their paper the following:
``Let us note that some corrections on the order of $\lambda^2 \rho^2\sim \Lambda r_r r_o$ to the lensing effects may exist, as well as other cosmological corrections in the general case of the FRW Universe. However, such ``short-distance" phenomena are perhaps too small to be of practical interest." In fact, using their notation where $\lambda\equiv\sqrt{\Lambda/3}$, the neglected term in question is $\Lambda r_g r_o/3$ and represents the contribution of $\Lambda$ to the squared invariant that was thrown away by assumption. 

Most recently, Kantowski, Chen and Dai \cite{Kantowski} aimed to reexamine all higher order corrections to the Einstein deflection angle including $\Lambda$-terms. They considered a Swiss-Cheese exact solution with Kottler in the vacuole and flat FLRW around it, and where angles are measured by comoving Friedmann observers. Kantowski et al. take careful account of the fact that the radius of the Kottler vacuole in the expanding FLRW universe is larger when the photon exits than when it entered. In fact, it is this expansion that in their approach gives rise to the most significant part of the correction term. In particular, the authors say, without the expansion there is no $\Lambda$-correction [see after their Eq. (23)]. And later, that it arises "since the cosmological constant contributes to the extra time ... the Schwarzschild mass has to act on the passing photons." Their lowest order correction in which $\Lambda$ appears is proportional to $m \sqrt{\Lambda}$, (their equation (32)) and they find that this could cause as much as a $0.02\%$ increase in the deflection angle of the light that passes through a rich cluster. It should be noted that a term like (33) can arise simply from aberration in an expanding universe; it does for example, when we apply the aberration factor $\sim (1+V)$ (mentioned at the end of section 2.1 above) to the mere Einstein bending angle $4m/R$ in a de Sitter universe, where $V=Hr$, $H=\sqrt{\Lambda/3}$. Also, Kantowski et al. in their introduction stress the desirability to produce formulas which, apart from $\Lambda$, contain only quantities that are measurably the same with and without $\Lambda$. "To conclude whether $\Lambda$ does or doesn't cause bending can easily depend on what is held in common and what property is compared in the two experiments" -- namely in a universe with and one without $\Lambda$. Now, their concluding formula (32) for the total deflection angle has the factor $m/r_0$ in front of it, $r_0$ being the coordinate radial distance of the point of closest approach. It could perhaps be argued that, in term of absolutes, $r_0$ might depend on the geometry of Kottler space and thus ultimately on $\Lambda$. In sum, Kantowski et al.'s work supports $\Lambda$-corrections to gravitational lensing in a flat $\Lambda$CDM model even if they arrive at different $\Lambda$-terms. 

Finally, Park \cite{Park} considered a customized McVittie metric and did an expansion of the null geodesic equations to first-order in the mass. Here also, second-order terms including $H^2=\Lambda/3$ terms were dropped out in the calculation in equations (11) and the final result (31), leading to the assertion that $\Lambda$ does not contribute to lensing except via the angular diameter distances. In fact, even at the same level of approximation used in his paper, $\Lambda$ contribution terms were simply omitted from the final lens equation (31) including a term with $H^2=\Lambda/3$ for reason of smallness, see footnote \footnote{We communicated our findings to the author of \cite{Park} and we are grateful to him for sending to us the following part that was omitted from his lens equation (31) under the assumption of smallness 
\be
\frac{2md_{SL}}{d_{S}d_{L}}\, \beta\, \left(- H^2 \frac{x_S d_L^2}{2} + ... \right).
\ee} 
Indeed, it seems fair to say that the omitted part with $H^2=\Lambda/3$ in the lens equation in \cite{Park} is in contradiction with the statement following equations (31-32) there and the conclusion presented in the paper. These terms are perhaps small for current observations but they do not cancel out as stated previously. 

\section{Conclusion}

Using a very simple example, we clarified that the effect of $\Lambda$ on the bending of light is physical and distinct from the abberation. 

We showed how the conclusions presented in \cite{SPF,Khriplovich,Park} about the non-contribution of $\Lambda$ to the bending of light and lensing phenomena are simply due to the fact that in the perturbative or expansion approaches used, the $\Lambda$ contribution terms were eliminated at some step or other of the calculation by too-stringent assumptions of smallness. 

A clarifying point is to realize that the Kottler vacuole surrounding the lens has a size that is larger than the lens and that when compared to the angular-diameter distances involved in the lens equation provides the derived contribution of $\Lambda$ to lensing. This point was also appreciated using a piecewise integration of the null geodesics in an Einstein-Strauss model \cite{Schucker1,Schucker2,Schucker3}. 

We also showed that the effect of $\Lambda$ can be ascribed to the invariant Weyl focusing. 

While the effect of $\Lambda$ on the light-bending and lensing is small as discussed in \cite{RindlerAndIshak2007,Ishaketal2008,Ishak2008}, the effect does not cancel out as has been claimed and is not a coordinate effect. 

\section*{Acknowledgments}
We thank Fergus Simpson, John Peacock, Alan Heavens, Minjoon Park, Thomas Schucker, and Mauro Sereno for stimulating discussions and correspondence. M.I. acknowledges that this material is based upon work supported in part by NASA under grant NNX09AJ55G and a grant from the Texas Space Grant Consortium. Part of the calculations for this work have been performed on the Cosmology Computer Cluster funded by the Hoblitzelle Foundation.


\label{lastpage}


\begin{thebibliography}{99}
%
%
\bibitem[Allen et al. 1998]{Allen} Allen S., M.N.R.A.S, \textbf{296}, 392, (1998).

\bibitem[Dodelson 2003]{Dodelson} Dodelson S., {\textit{Modern Cosmology}}, Academic Press (2003).
%

\bibitem[Dyer 1977]{Dyer1977} Dyer C., Mon. Not. Roy. Astro. Soc. {\textbf{180}}, 231 (1977).

\bibitem[Finelli et al. 2007]{Finelli} Finelli F., M. Galaverni, A. Gruppuso, Phys.Rev. D {\textbf{75}}, 043003 (2007).

\bibitem[Ishak 2008]{Ishak2008} Ishak M., Phys. Rev. D \textbf{78}, 103006, (2008).

\bibitem[Ishak et al. 2008]{Ishaketal2008} Ishak M., W. Rindler, J. Dossett, J. Moldenhauer, C. Allison, Monthly Notices of the Royal Astronomical Society, Volume \textbf{388}, Issue 3, p1279 (2008).
%
\bibitem[Kagramanova et al. 2006]{Kagramanova}  Kagramanova V., J. Kunz, C. Lammerzahl, Phys.Lett. B {\textbf{634}} 465-470 (2006).

\bibitem[Kantowski et al. 2009]{Kantowski} Kantowski R., Chen B., Dai X, arXiv:0909.3308  (2009), ApJ submitted.

\bibitem[Khriplovich et al. 2008]{Khriplovich} Khriplovich I., A. Pomeransky, Int. J. Mod. Phys. D \textbf{17}, 2255-2259 (2008).

\bibitem[Kottler 1918]{Kottler} Kottler F., Ann. Phys. {\textbf{361}}, 401 (1918).

\bibitem[Lake 2007]{Lake2007} Lake K., arXiv:0711.0673.

\bibitem[Miraghaei et al. 2008]{Miraghaei} Miraghaei H., M. Nouri-Zonoz, arXiv:0810.2006v1 [gr-qc] (2008).

\bibitem[Park 2008]{Park} Park M.,   Phys. Rev. D.\textbf{78}, 023014 (2008).  

\bibitem[Rindler 2006]{Rindler} Rindler W., \textit{Relativity: Special, General, and Cosmological, Second Edition} (Oxford University Press, 2006).

\bibitem[Rindler \& Ishak 2007]{RindlerAndIshak2007} Rindler W., M. Ishak, Phys. Rev. D \textbf{76} 043006 (2007).

\bibitem[Sachs et al. 1961]{Sachs1961}Sachs R.K., Proc. Roy. Soc. London. A {\textbf{264}}, 309 (1961).

\bibitem[Schneider et al. 1992]{Schneider1992} Schneider P., J. Ehlers and E.E. Falco, \textit{Gravitational Lenses} (Springer-Verlag, 1992).

\bibitem[Schucker 2009]{Schucker1} Schucker T., General Relativity and Gravitation, \textbf{41}:67-75,(2009). 

\bibitem[Schucker 2008]{Schucker2} Schucker T., N. Zaimen, Astrono. \& Astrophys., \textbf{484}, 103 (2008).

\bibitem[Schucker 2009]{Schucker3} Schucker T., General Relativity and Gravitation \textbf{41}:1595-1610  (2009).  

\bibitem[Sereno 2008]{Sereno2007} Sereno M., Phys.Rev. D {\textbf{77}}, 043004 (2008).

\bibitem[Simpson et al 2008]{SPF} Simpson F., J. A. Peacock, A. F. Heavens, arXiv:0809.1819v1 [astro-ph] (2008)

\bibitem[Smail et al. 1991]{Smail1991} Smail I. et al., M.N.R.A.S, \textbf{252}, 19 (1991).

\bibitem[Wald 1984]{Wald1984} Wald R. \textit{General Relativity} (University of Chicago Press, 1984).

\end{thebibliography}
\end{document}